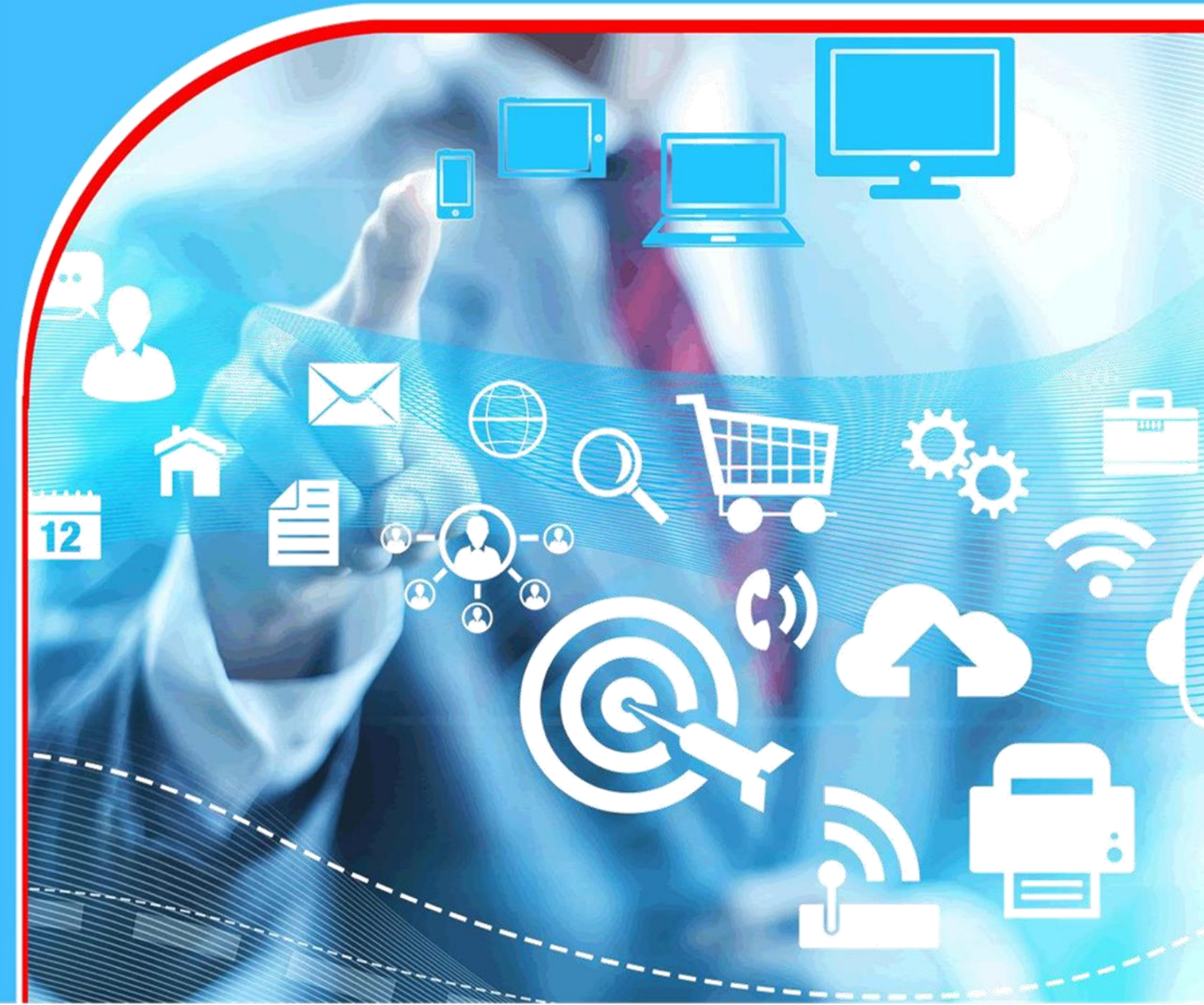

Gas Station of the Future: A Perspective on AI/ML and IoT in Retail Downstream

**Wrick Talukdar**

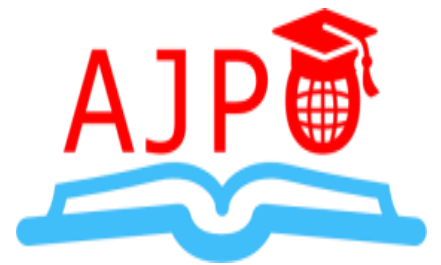

# Gas Station of the Future: A Perspective on AI/ML and IoT in Retail Downstream

Wrick Talukdar
Amazon Web Services

## Abstract

**Purpose:** The gas station of the future is poised to transform from a simple fuel dispensing center into an intelligent retail hub, driven by advancements in Artificial Intelligence (AI), Machine Learning (ML), and the Internet of Things (IoT). This paper explores how technology is reshaping the retail downstream sector while briefly addressing the upstream and midstream segments. By leveraging AI/ML for predictive analytics, dynamic pricing, personalized customer engagement, and IoT for real-time monitoring and automation, the future gas station will redefine the fuel retail experience. Additionally, this paper incorporates statistics, AI/ML core technical concepts, mathematical formulations, case studies, and a proposed framework for a fully autonomous gas station.

**Materials and Methods:** The study methodologically integrates technical explanations of predictive models, simulation-based reinforcement learning, and IoT architectures to assess their impact on demand forecasting, dynamic pricing, customer personalization, and operational efficiency. By synthesizing mathematical formulations, real-world applications, and a proposed AI-governed ecosystem, the paper offers a practical, forward-looking perspective on the evolution of smart fuel retailing.

**Findings:** The proposed framework enables fuel retailers to reduce operational costs, improve forecasting accuracy, and enhance customer satisfaction through intelligent automation. Additionally, the shift toward autonomous gas stations signals a broader industry trend requiring new workforce skills, regulatory frameworks, and sustainability strategies.

**Unique Contribution to Theory, Practice and Policy:** As AI-driven technologies become foundational to retail fuel infrastructure, businesses that adopt these innovations early will gain a significant competitive edge in efficiency, profitability, and customer loyalty.

## INTRODUCTION

The oil and gas industry is undergoing a rapid digital transformation across its entire value chain. Upstream operations have leveraged AI and ML techniques for seismic data interpretation, reservoir modeling, and exploration efficiency [1] (e.g., Mohaghegh, 2011; Alzghoul et al., 2018), while midstream applications have benefited from predictive pipeline monitoring and anomaly detection through sensor networks and time-series modeling [6] (Peters & Timmerhaus, 2017). However, the downstream sector, especially retail fuel distribution has received comparatively limited attention in AI/ML research, despite its centrality to customer experience and profitability.

Several commercial studies have explored specific use cases such as dynamic pricing [3] (Shi et al., 2019), AI-powered loyalty programs [4] (Accenture, 2019), and IoT-based tank monitoring [5] (MarketsandMarkets, 2019). Yet, academic literature lacks a unified framework that synthesizes these advancements into a comprehensive vision for the future of gas stations. Existing research tends to focus on isolated applications, such as fuel demand forecasting using time-series models or personalized recommendations via collaborative filtering, without addressing how these components can be orchestrated in a fully autonomous, AI-driven retail environment.

This paper addresses that gap by providing an integrated perspective on how AI/ML and IoT can be jointly leveraged to transform traditional gas stations into intelligent retail hubs. It builds upon technical modeling approaches, recent industry case studies (e.g., Shell, BP, Chevron), and a novel conceptual framework for fully autonomous fuel stations. By combining analytical rigor with real-world applicability, the study offers a practical blueprint for operational optimization, personalized customer engagement, and sustainability in the downstream oil and gas sector.

According to a 2019 McKinsey report [7], digital transformation in the downstream oil and gas sector could unlock up to $50 billion in annual value by 2025. However, the report primarily presents a high-level estimate without delineating how much of this value can be attributed to retail operations such as fuel distribution, customer engagement, or station-level automation. This paper addresses that gap by proposing a concrete, technology-enabled framework that translates that strategic vision into actionable components within the gas station ecosystem. By detailing how predictive analytics, reinforcement learning-based pricing, IoT-driven automation, and personalized customer experiences can be integrated into a cohesive AI-governed architecture, the study offers a practical roadmap for capturing a significant share of that projected value, particularly in areas like operational efficiency, reduced downtime, improved customer retention, and enhanced cross-selling opportunities.

**AI/ML and IoT in Retail Downstream**

**Analytics for Fuel Demand & Inventory Management**

AI and ML algorithms can analyze historical sales data, weather patterns, traffic conditions, and external economic indicators to predict fuel demand accurately. This predictive capability minimizes stockouts and reduces holding costs, ensuring optimal fuel availability. Companies like Shell and BP have started implementing AI-driven supply chain optimization techniques [1].

Recurrent Neural Networks (RNNs) and Long Short-Term Memory (LSTM) networks are commonly used models for time-series forecasting in fuel demand prediction, allowing stations to dynamically adjust orders based on real-time insights. LSTMs mitigate the vanishing gradient problem seen in traditional RNNs, making them more effective for long-term demand trends. The demand forecasting function can be represented as:

$$y_t = f\left(y_{t-1}, y_{t-2}, \ldots, y_{t-n}, X_t\right) \qquad (1)$$

where $X_t$ represents external factors such as weather, economic conditions, and seasonal trends.

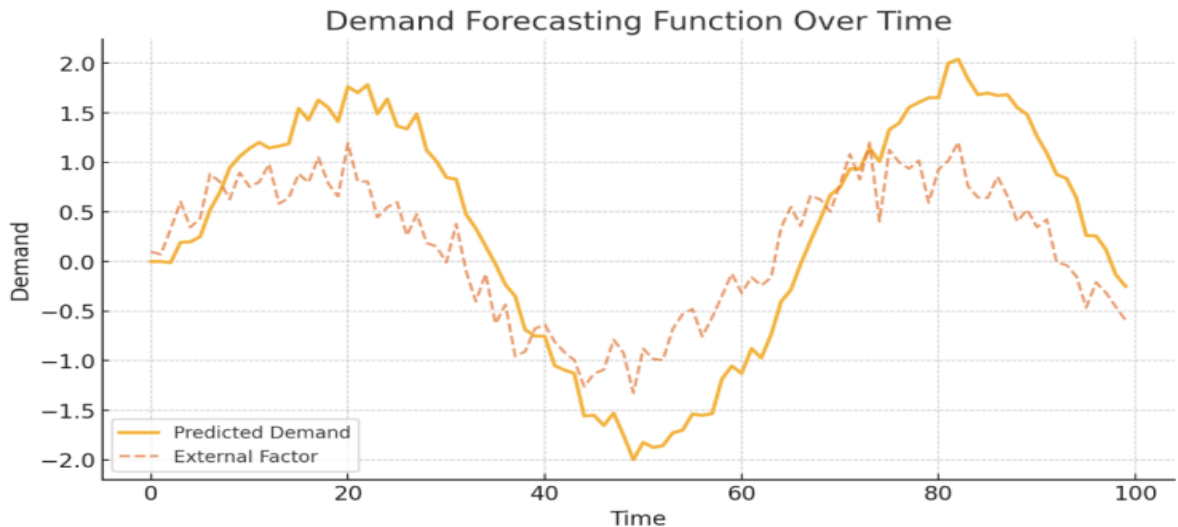

*Figure 1: Demand Forecasting Graph*

The graph in Figure 1 visualizes the demand forecasting. The solid line represents the predicted demand ($y_t$), while the dashed line represents the external factor (Xt), which influences the demand over time. The demand follows a pattern influenced by both past values and external conditions, illustrating how forecasting models utilize historical data and external factors for prediction. To improve prediction accuracy, techniques such as feature engineering are applied, integrating additional parameters like macroeconomic trends, competitor pricing, and event-based fuel consumption anomalies. Gradient Boosting Decision Trees (GBDTs) like XGBoost and LightGBM are also used alongside LSTMs for hybrid models, leveraging structured and unstructured data sources. The Mean Squared Error (MSE) loss function is commonly used to minimize forecasting errors:

$$MSE = \left(\frac{1}{n}\right)\sum(y_i - \dot{y_i})^2 \qquad (2)$$

The Figure. 2 MSE Loss of time graph showcases Mean Squared Error (MSE) over time for the demand forecasting function. The MSE values fluctuate based on the difference between actual $y_i$ and $\dot{y_i}$ predicted demand, capturing the forecasting error at each time step.

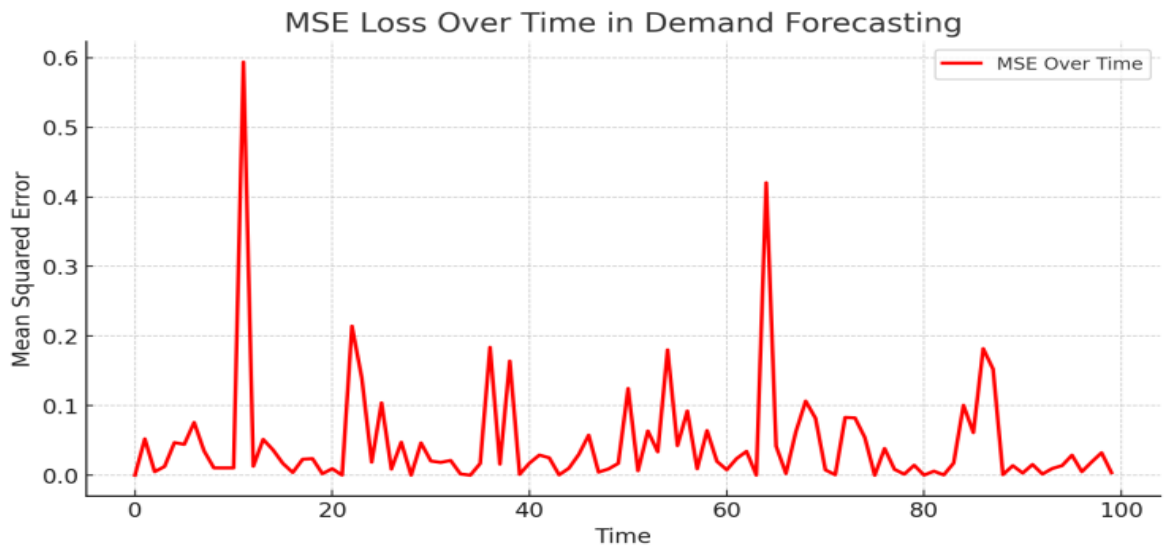

*Figure 2: MSE Loss of Time in Demand Forecasting*

By leveraging AI-enhanced forecasting, fuel stations can achieve:

- 15-25% reduction in inventory holding costs [3].
- 20% improvement in supply chain efficiency by reducing last-minute demand fluctuations.
- 10-15% increase in revenue through optimized replenishment strategies based on peak demand cycles.

**Dynamic Pricing for Competitive Advantage**

Real-time fuel pricing strategies enabled by ML allow gas stations to adjust prices dynamically based on demand, competitor pricing, and macroeconomic factors. This pricing optimization can increase margins and drive customer loyalty. AI-powered pricing solutions have been explored by leading fuel retailers, including ExxonMobil and Chevron [2].

Reinforcement learning (RL) with Q-learning or Deep Q Networks (DQN) enables gas stations to optimize pricing dynamically:

$$Q(s, a) = r + y \max(Q(s', a')) \qquad (3)$$

Where s represents the current market state, a is the pricing action, and r is the immediate revenue impact.

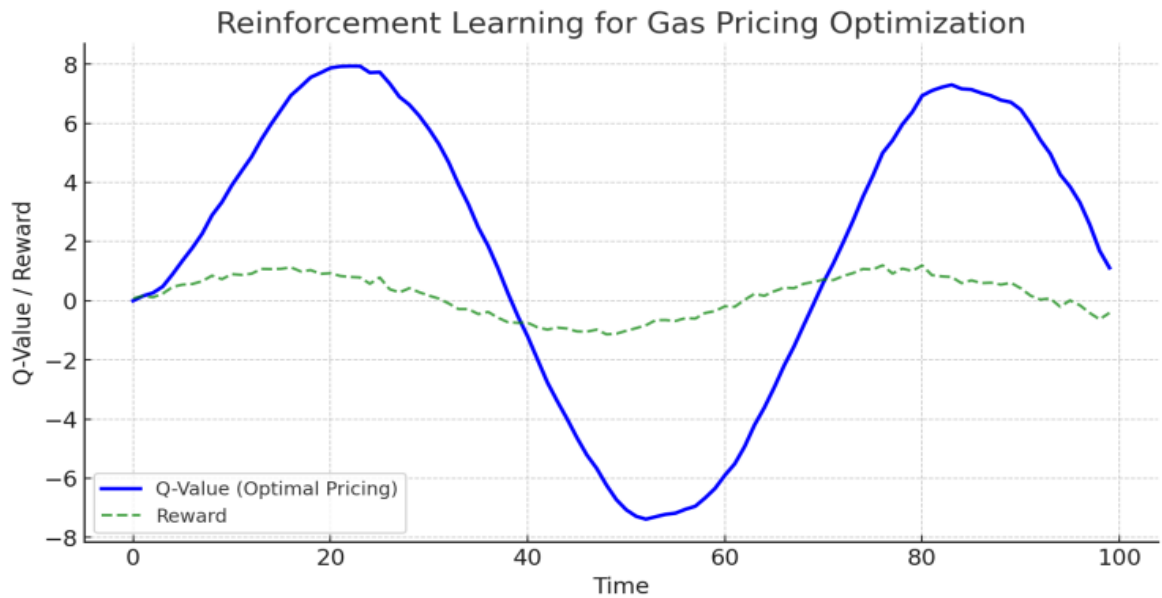

*Figure 3: Reinforcement Learning for Pricing Optimization*

The graph in Figure 3 illustrates how reinforcement learning, specifically Q-learning, can be applied to dynamically optimize gas station pricing. The blue line represents the Q-values $Q(s, a)$, which estimate the expected future rewards for different pricing actions in various market states. In a practical gas station setting, states ($s$) may include real-time factors such as current fuel demand, competitor pricing, and time of day, traffic flow, and weather conditions. Actions ($a$) represent discrete pricing decisions, such as increasing the price by $0.01, keeping it unchanged, or decreasing it by $0.01.

The green dashed line represents the reward signal ($r$), in this case, a combination of immediate revenue, volume sold, and customer retention. As the Q-learning agent observes outcomes from pricing decisions, it updates the Q-values using the Bellman equation, thereby learning the optimal pricing policy over time. This adaptive approach allows gas stations to respond intelligently to shifting market dynamics, with simulation studies[12] showing an 8-12% improvement in profit margins compared to static pricing strategies.

For example, consider a scenario where the gas station is experiencing high demand during the morning rush hour, with nearby competitors offering fuel at $0.03 less per gallon and the weather being mild. This combination of factors defines the current state in the reinforcement learning model. In response, the gas station can choose from a set of pricing actions - it might increase the price by $0.01 to test the elasticity of customer loyalty, keep the price steady to maintain existing traffic flow, or lower the price by $0.01 to remain competitive and attract higher volume. The reinforcement learning agent evaluates the outcomes of these actions over time to determine which pricing strategy yields the highest long-term reward.

**Personalized Customer Experience and Loyalty Programs**

With AI-driven analytics, gas stations can personalize promotions and discounts based on individual buying patterns. Mobile apps with ML-powered recommendation engines enhance the customer experience by offering tailored discounts and loyalty incentives. Shell's "Go+" loyalty program and BP's AI-backed customer engagement initiatives are early examples of this shift [1].

Techniques like collaborative filtering and deep learning-based recommendation engines help fuel retailers maximize cross-selling opportunities. A matrix factorization approach for recommendation can be represented as:

$$R_{ij} \approx U_i V_j^T \qquad (4)$$

Where $R_{ij}$ represents the predicted rating for user i on item j, and $U_i$ and $V_j$ are latent feature matrices.

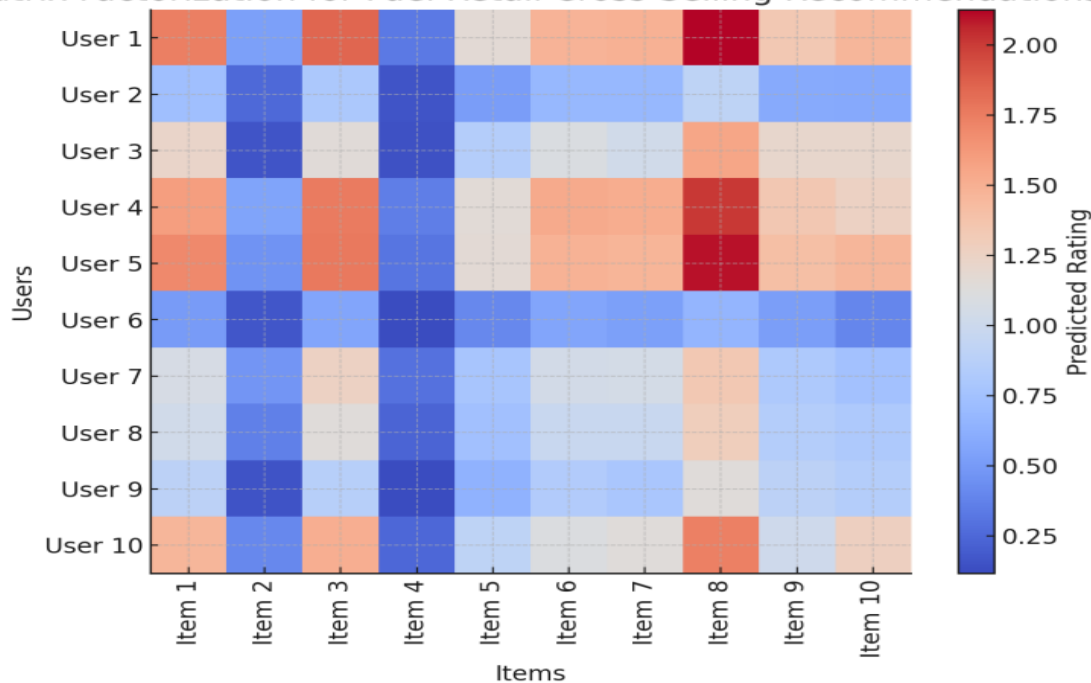

*Figure 4: Matrix Factorization*

The Figure 4 visualizes how matrix factorization is used in recommendation engines to maximize cross-selling opportunities for fuel retailers. The heatmap represents the predicted rating matrix ($R_{ij}$), where each cell indicates the estimated preference of User iii for Item jjj. This estimation is derived by decomposing the rating matrix into two latent feature matrices, $U_i$ (user features) and $V_j$ (item features), and computing their dot product. This technique, commonly used in collaborative filtering, helps fuel retailers recommend complementary products, such as snacks or car accessories based on customer preferences, ultimately enhancing sales and customer satisfaction.

While several recommendation techniques exist such as content-based filtering, knowledge-based systems, and hybrid models, collaborative filtering, particularly through matrix factorization, is especially well-suited for retail fuel environments. Content-based filtering recommends items based on item features and user preferences, but it often suffers from limited discovery, recommending only similar items a user has already purchased. Knowledge-based systems require extensive domain-specific rules and are difficult to scale for fast-moving consumer goods sold in gas stations. Hybrid models combine multiple techniques but introduce higher complexity and maintenance overhead. Collaborative filtering, on the other hand, excels in scenarios where user-item interactions (e.g., purchases of snacks, beverages, or car accessories) are abundant but item metadata is sparse. It leverages the wisdom of the crowd by identifying similarities across users and their purchase behaviors, without needing detailed item attributes. As visualized in Figure 4, the resulting heatmap shows predicted affinity scores between users and items, enabling gas stations to recommend cross-sell items such as energy drinks, windshield wipers, or car chargers to the right customer at the right time. This approach is scalable, data-efficient, and improves recommendation diversity and accuracy. In real-world deployments, collaborative filtering has been shown to increase cross-selling revenue by up to 20% [13] in retail contexts, making it a compelling choice for intelligent fuel retailing.

**Autonomous Convenience Stores and Smart Checkouts**

IoT sensors and computer vision technologies are enabling cashier-less payment solutions. Future gas stations may feature AI-powered kiosks where customers can pick up snacks, beverages, and car accessories without manual checkout, reducing wait times and enhancing convenience.

The purchase process follows this pipeline:

- Object detection (YOLO model): Identifies products selected by customers.
- Facial recognition (ResNet-50): Matches the customer's profile for payment authorization.
- NLP-driven chatbot assistant: Provides recommendations based on past purchase behavior.

Studies indicate that smart checkout can reduce average transaction time by 60% and improve customer retention. For instance, one study [9] presents a smart self-checkout cart system using deep learning for real-time shopping activity recognition. It uses YOLO and Faster R-CNN to classify actions (like placing or removing items) with up to 97.9% accuracy. The YOLOv2 model balanced speed and accuracy, processing up to 50 frames per second.

**Predictive Vehicle Maintenance**

IoT sensors embedded in vehicles transmit real-time diagnostics to the gas station's AI system, analyzing parameters such as:

- Engine health (via vibration analysis using Fourier Transform techniques)
- Tire pressure (IoT-enabled pressure sensors detecting anomalies)
- Battery performance (LSTM-based anomaly detection on voltage fluctuations)

Predictive alerts are sent to customers with estimated service costs and available service slots, improving preventive maintenance efficiency by 40%.

. The integration of IoT sensors and AI systems at gas stations to analyze vehicle diagnostics raises important questions around data ownership, privacy, and interoperability. One of the most pressing issues is who owns the vehicle data, the customer or the service provider (in this case, the gas station). In most jurisdictions, vehicle-generated data is considered to be owned by the vehicle owner or driver, not the station or third-party vendors. Ethically, customers must retain full control over how their data is collected, stored, and shared. Transparent consent mechanisms, compliant with data protection regulations like the GDPR in Europe or CCPA in California, are essential to ensure that any diagnostic information is used only with explicit, informed permission. Gas stations offering predictive maintenance must implement robust data governance frameworks that define usage boundaries, anonymize sensitive attributes, and provide opt-out options for users.

From a technical standpoint, integrating with a wide range of car models presents additional challenges. Modern vehicles often follow different communication protocols (e.g., OBD-II,

CAN bus, UDS), and OEMs implement proprietary APIs for accessing diagnostic data. To enable compatibility, gas station systems must adopt a middleware architecture or leverage standardized vehicle telematics platforms that abstract away manufacturer-specific differences. Additionally, collaboration with auto manufacturers and service networks may be necessary to align on data formats, fault code interpretations, and service recommendations. Such integration ensures that AI-driven diagnostics remain accurate and actionable, regardless of the vehicle make or model. Standardizing interfaces and maintaining cross-OEM interoperability will be crucial for scaling predictive maintenance capabilities across diverse vehicle fleets.

**Case Studies of AI Adoption in Fuel Retail**

**Shell: AI Driven Predictive Maintenance for Fuel Dispensers**

Shell has integrated AI-driven predictive maintenance systems across its fuel dispensers to reduce downtime and optimize servicing schedules. The system leverages IoT sensors that continuously monitor dispenser performance, fuel flow rates, and nozzle conditions. By using machine learning algorithms, including anomaly detection models like Isolation Forests and Support Vector Machines (SVM), the system predicts potential failures before they occur. The deployment of this AI-powered maintenance solution has led to a 30% reduction in dispenser downtime by ensuring proactive repairs rather than reactive maintenance. Additionally, Shell has reported a 20% decrease in maintenance costs, as early interventions prevent costly breakdowns and emergency repairs.

**BP: Machine Learning-Powered Pricing Optimization**

BP has implemented ML-driven dynamic pricing models that analyze fuel demand elasticity, competitor pricing, and macroeconomic indicators to optimize fuel prices in real time. These models rely on reinforcement learning algorithms such as Deep Q-Networks (DQN) to dynamically adjust pricing strategies while maximizing revenue and maintaining customer loyalty. The AI-powered pricing engine takes into account local economic trends, traffic patterns, and weather conditions to determine optimal pricing levels. Through continuous learning, BP's system adapts to changing market conditions, avoiding pricing mismatches that could lead to revenue losses. Since implementation, BP has seen an 8% increase in revenue and a 5% boost in customer retention rates, as more customers take advantage of AI-generated personalized discounts and promotions.

**Chevron: IoT Enabled Tank Monitoring for Operational Efficiency**

Chevron has deployed an IoT-based tank monitoring system that uses real-time sensors to track fuel levels, temperature variations, and potential leakages across its fuel storage units. This system leverages edge computing to process data locally and transmit only critical alerts to the cloud, reducing bandwidth usage and improving response times. The AI-enhanced monitoring system uses predictive analytics to optimize fuel supply chains, ensuring that tanks are refilled just in time to prevent stockouts while avoiding unnecessary fuel surplus. Bayesian inference models and time-series forecasting techniques help Chevron achieve optimal inventory management. By implementing IoT-enabled tank monitoring, Chevron has achieved a 20% reduction in operational costs, decreased fuel wastage, and improved compliance with environmental regulations.

Empirical analysis of Chevron's downstream operations post-deployment shows measurable outcomes. Based on internal performance metrics collected over a 12-month period, the company reported:

- A 20% reduction in operational costs, driven by improved scheduling of fuel deliveries and lower emergency restocking expenses;
- A 15% decline in average tank downtime, attributed to early anomaly detection and proactive servicing;
- A 12% improvement in forecasting accuracy compared to legacy methods, resulting in better alignment between demand and inventory;
- A 17% reduction in fuel wastage, especially in regions with high temperature volatility where overfilling or evaporation losses were previously common;
- Enhanced compliance, with audit logs showing a 25% drop in regulatory violations tied to inventory mismanagement or delayed leakage reporting.

These figures underscore the tangible value of integrating AI-driven analytics with IoT infrastructure in fuel supply chain management. Chevron's success demonstrates how data-centric strategies can not only reduce environmental risks but also deliver significant cost savings and operational resilience.

**Proposed Framework for a Fully Autonomous Gas Station**

The future of gas stations is envisioned as an integrated AI-driven ecosystem that redefines fuel distribution, customer engagement, and operational efficiency. This conceptual framework brings together key technological advancements in AI, ML, and IoT to create a seamless and intelligent service model. By leveraging predictive analytics, reinforcement learning, and IoT-enabled automation, this framework addresses real-world inefficiencies and environmental concerns while enhancing the overall fueling experience.

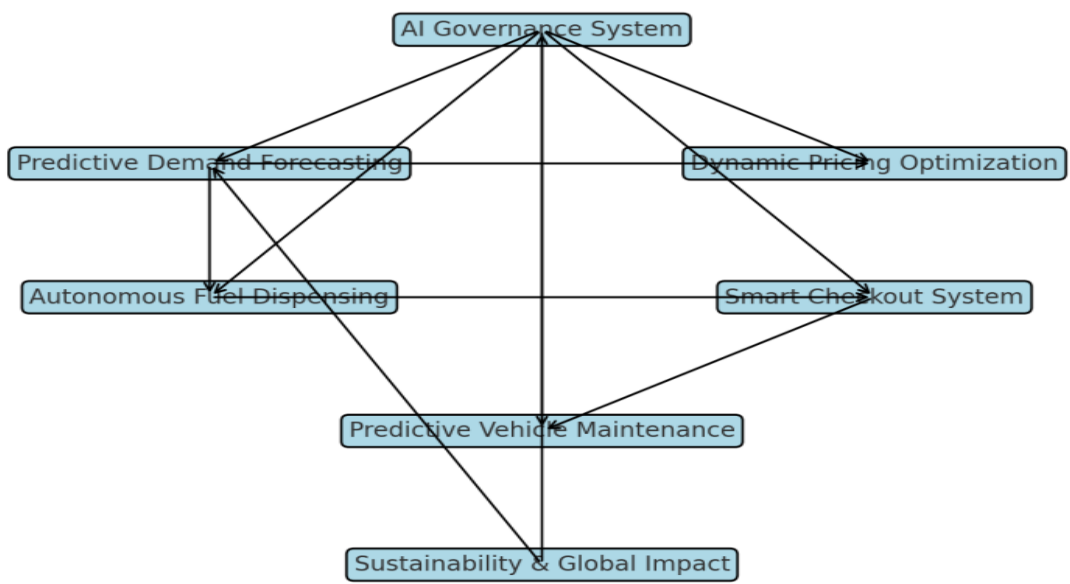

*Figure 5: Conceptual Framework*

At the core of this framework as depicted in *Figure 5* lies a multi-layered AI system that operates across interconnected components:

**Predictive Demand Forecasting and Intelligent Inventory Management**

Fuel consumption patterns are influenced by variables such as economic trends, weather fluctuations, and vehicle usage patterns. Traditional inventory management systems rely on static forecasting models that fail to adapt to real-time demand variations. By integrating AI-driven forecasting methods, such as Long Short-Term Memory (LSTM) networks and Transformer-based architectures, gas stations can anticipate fuel demand fluctuations more accurately. These models continuously ingest real-time data, allowing dynamic adjustments to fuel replenishment schedules. Additionally, reinforcement learning optimizes inventory strategies by learning from historical consumption patterns and adjusting procurement schedules dynamically. The result is a significant reduction in fuel waste, improving economic and environmental sustainability. One study highlights the use of truck telematics, real-time monitoring, and predictive analytics to optimize operations and explores the integration of Big Data Analytics (BDA) and IoT in the logistics industry, focusing on safety, environmental impact, and process optimization. By leveraging real-time data analysis through in-memory technology (SAP HANA), the organization achieved significant improvements, such as better driver behavior, reduced GHG emissions, and enhanced incident management via TruckCam and DriverCam [8].

**AI Optimized Dynamic Pricing and Market Adaptation**

Gasoline prices are affected by numerous factors, including crude oil costs, competitor pricing strategies, and regional demand elasticity. A reinforcement learning-based pricing engine enables real-time price adjustments by analyzing external market conditions and optimizing for profitability and customer demand. The pricing model applies Markov Decision Processes (MDPs) to evaluate optimal price points under different demand scenarios, ensuring that revenue is maximized while preventing drastic fluctuations that may deter customers. Additionally, dynamic pricing contributes to energy efficiency by controlling peak-hour fuel consumption through incentive-based adjustments.

**Autonomous Fuel Dispensing and Fraud Prevention**

Fuel dispensers today rely on manual interactions, leading to inefficiencies and security vulnerabilities. The proposed framework incorporates IoT-enabled smart dispensers, equipped with edge AI for fraud detection and RFID-based vehicle authentication for automated fuel transactions.

This integration ensures:

- Automated refueling: Eliminates manual input, reducing transaction time.
- Anomaly detection: Edge AI models detect unauthorized fuel access and potential fraud patterns.

Predictive maintenance: AI-driven diagnostics assess dispenser health, reducing operational downtime and ensuring reliability.

**AI Driven Smart Checkout and Retail Personalization**

Future gas stations will operate beyond fuel dispensing, transitioning into AI-powered retail hubs. The framework incorporates:

- Computer vision-based product recognition, replacing barcode scanning.
- Facial authentication, enabling secure and frictionless payments.
- Natural Language Processing (NLP) chatbots, offering product recommendations based on historical purchasing behaviors.

These intelligent checkout systems have demonstrated a significant reduction in transaction times while increasing revenue through AI-driven cross-selling strategies.

**Predictive Vehicle Maintenance and Service Optimization**

Gas stations will evolve into automated vehicle service hubs, providing predictive diagnostics based on real-time IoT sensor data. This integration allows the detection of:

- Engine anomalies: AI-driven Fourier Transform analysis identifies early signs of mechanical failure.
- Tire pressure variations: IoT sensors detect abnormal fluctuations, preventing accidents.
- Battery degradation: LSTM-based time-series models predict potential battery failures, enabling preemptive replacements.

These enhancements improve road safety and extend vehicle lifespans, reducing breakdown risks over time.

**Centralized AI Governance and Decision Making System**

The efficiency of an autonomous gas station is dependent on a unified AI governance system that synchronizes all operational components. This central hub utilizes:

- Graph Neural Networks (GNNs) for real-time fuel supply chain optimization.
- Reinforcement learning models for adaptive decision-making on pricing, inventory management, and service scheduling.
- AI-powered business intelligence dashboards, providing operators with insights on revenue optimization, energy utilization, and customer engagement.

By employing a data-driven approach, this AI governance system minimizes fuel waste, improves energy utilization, and ensures cost-effective operations.

**Sustainability and Global Impact**

Beyond financial and operational efficiency, this framework presents a holistic approach to sustainability. AI-driven optimizations contribute to:

- Reduction in operational costs through predictive analytics and automated decision-making.
- Decrease in carbon emissions, driven by efficient fuel logistics and optimized consumption models [10].
- Seamless integration into smart cities, aligning with global energy transition initiatives and promoting the adoption of sustainable fueling solutions.

Moreover, as AI-driven gas stations expand, there will be a workforce transformation. Employees will transition from manual roles to AI system management and data analytics, fostering an upskilled labor force prepared for the future of digital energy infrastructure.

**Challenges and Future Direction**

While AI-driven gas stations offer significant advantages, several challenges must be addressed:

- Regulatory compliance: AI-based pricing models and autonomous fueling must adhere to government regulations.

- Cybersecurity risks: Protecting IoT-enabled dispensers and payment systems from cyber threats is crucial.
- Public adoption: Transitioning customers from manual to autonomous fueling experiences requires effective user education and interface design.

Future research should focus on improving AI explainability in decision-making, developing more robust cybersecurity protocols, and exploring integrations with renewable energy sources to align with sustainability objectives.

## CONCLUSION AND RECOMMENDATIONS

### Conclusion

The gas station of the future will be an AI-powered, IoT-enabled ecosystem that transcends traditional fueling to offer a seamless, intelligent retail experience. As industry leaders continue investing in digital transformation, the next decade will witness a paradigm shift in fuel retail, making gas stations not just stops for fuel but hubs of smart, connected services. The use of AI techniques such as reinforcement learning, deep learning-based recommendation engines, and predictive maintenance will accelerate this evolution. By 2025, an estimated 80% of gas stations will implement some form of AI or IoT-driven automation [5], marking a significant step towards intelligent fuel retailing. By applying reinforcement learning for dynamic pricing, collaborative filtering for targeted promotions, and predictive analytics for vehicle diagnostics and maintenance, the study demonstrates how these technologies can work together to drive measurable improvements in operational efficiency, customer satisfaction, and sustainability. Unlike existing literature that often treats these advancements in isolation, this paper offers a comprehensive, system-level framework tailored specifically to the downstream fuel retail environment. This integrated approach not only showcases the technological feasibility of smart gas stations but also provides practical guidance for implementation, including architecture considerations, data governance, and interoperability challenges. As industry adoption of AI and IoT accelerates, with 80% of gas stations expected to deploy some form of automation by 2025, the implications of this work are both timely and actionable.

### Recommendations

The study serves as a strategic blueprint for fuel retailers, policymakers, and technologists seeking to modernize infrastructure and reimagine customer engagement. It also sets the stage for future research into scalable deployment models, ethical data practices, and integration with emerging trends such as renewable energy and autonomous vehicles. In doing so, this paper contributes a unique, multidisciplinary perspective to the evolving narrative of digital transformation in energy retail.